\documentclass{PoS}

\usepackage{siunitx}
\usepackage{amsmath}
\usepackage{floatrow}
\usepackage{subfig}

\title{Seasonal Variation of Atmospheric Neutrinos in IceCube}

\ShortTitle{Seasonal Variation of Atmospheric Neutrinos}

\author{
The IceCube Collaboration\footnote{For collaboration list, see PoS(ICRC2019) 1177.}\\
{\itshape \href{http://icecube.wisc.edu/collaboration/authors/icrc19_icecube}{http://icecube.wisc.edu/collaboration/authors/icrc19\_icecube}}\\
E-mail: \email{wiebusch@physik.rwth-aachen.de}
}

\abstract{
The IceCube Neutrino Observatory detects atmospheric muon neutrinos above 100 GeV at a rate of about 100 000 per year. These neutrinos originate from decays of charged pions and kaons in cosmic ray air showers. Their flux depends on the probability of production and decay of the parent mesons, and is thus sensitive to the stratospheric temperature.
Neutrino rates from 8 years of operation of the detector are correlated with the atmospheric temperature profile as measured by the Atmospheric Infrared Sounder (AIRS). An analysis of this correlation provides a test of models of hadronic interactions in atmospheric air showers. This analysis of neutrinos complements the analysis of the correlation of atmospheric muons with temperature that is presented in another paper at this conference.

\vspace{4mm}
{\bfseries Corresponding authors:}
 Patrick Heix$^{2}$, \speaker{Serap Tilav}$^{1}$,
 Christopher Wiebusch$^{2}$, Marit Z\"ocklein$^{2}$
\\
{$^{1}$ \itshape Bartol Research Institute and Dept. of Physics and Astronomy, University of Delaware, Newark, DE 19716, USA}\\
{$^{2}$ \itshape III. Physikalisches Institut, RWTH Aachen University, D-52056 Aachen, Germany}\\
}

\FullConference{36th International Cosmic Ray Conference -ICRC2019-\\
		July 24th - August 1st, 2019\\
		Madison, WI, U.S.A.}

\begin{document}

\section{Introduction\label{sec:intro}}

Besides its primary aim of studying cosmic neutrinos, IceCube also measures neutrinos and muons produced by the interaction of cosmic rays with the Earth's atmosphere with unprecedented statistics.
The flux of atmospheric neutrinos, as well as muons, is interesting, because it is produced by weak decays of unstable mesons. Accordingly, it directly probes the particle physics of hadronic air showers.
The flux of muon neutrinos can be approximated  \cite{Gaisser:1990vg}  by integrating the production yield of neutrinos over the atmospheric slant depth $X$:
\begin{equation}
    % see https://wiki.icecube.wisc.edu/index.php/Prediction_of_seasonal_variations_of_atmospheric_neutrinos
    \label{eq:prodyield}
    \phi_\nu(E_\nu,\theta) = \phi_N(E_\nu) \cdot  \int_0^{X_\mathrm{\footnotesize{ground}}}
    \left (
\frac{A_{\pi\to\nu}(X)}{1+B_{\pi\to \nu}(X) \cdot{ E_\nu \cos{(\theta^*)}\over \epsilon_\pi(T(X)) }}
+
\frac{A_{K\to\nu}(X)}{1+B_{K\to \nu}(X) \cdot{ E_\nu \cos{(\theta^*)}\over \epsilon_K(T(X)) }}
\right ) \, dX
\end{equation}
where $\phi_N(E_\nu)$  is the primary spectrum of cosmic ray nucleons (N)
evaluated at the energy of the neutrino.
The two terms in the parentheses are the production yield  $P(E_\nu,\theta,X) \equiv (\dots)$ of neutrinos from parent pions and kaons respectively as a function of atmospheric depth and direction.
The numerators $A_{i\to\nu}$ include the production yield, and branching ratios for the decay, with kinematic factors for the  respective parent mesons.
The denominators reflect  the competition between decay and interaction of
secondary mesons in the atmosphere.
The factors $B_{i\to\nu}$ include interaction cross sections of mesons and nucleons with kinematic factors, 
$\theta^* (\theta)$ is the neutrino zenith angle at its point of production in the atmosphere that is given by the observed zenith $\theta$ taking into account the curvature of the Earth, and $\epsilon_i (T)$ are characteristic energies that reflect whether re-interaction or decay of the parent mesons dominates.
These characteristic energies 
depend on the local air density which results in a linear dependence on the temperature $T$:
\begin{equation} \label{eq:epsilon}
\epsilon_i 
%=  \frac{c \cdot m_{i} \cdot h_0}{ \tau_{i}}
= T(X) \cdot \frac{R}{Mg}
\frac{c \cdot m_{i} }{ \tau_{i}}
\end{equation}
where $R$, $M$, and $g$ are the ideal gas constant, molar mass, and gravitational acceleration, $c$ the speed of light, and  $m_i$ and $\tau_i$ are the mass and decay time of the mesons.
Typical values of the characteristic energies are $\epsilon_\pi \approx \SI{125}{GeV} $ and $\epsilon_K \approx \SI{850}{GeV} $.
At energies  $E_\nu \cdot \cos{(\theta^*)} $ less than $ \epsilon_i $, most mesons decay, and the neutrino flux is independent of the atmospheric temperature. It simply follows the spectrum of parent cosmic rays.
For energies above $ \epsilon_i $, %the critical values, 
decays become increasingly disfavored and the resulting flux correlates %linearly 
with the atmospheric temperature, and has a spectrum one power steeper than that of primary cosmic rays.
As the correlation of atmospheric neutrino flux with the atmospheric temperature depends on the relative contribution from the pion and kaon parent particles, the measurement is sensitive to the hadronic physics of air showers.
 
The  total  rate of neutrino events $R_\nu$ is correlated with  the effective atmospheric temperature $T_{eff} $ as:
  \begin{equation} \label{eq:corr}
     \frac{R_\nu (t) - \langle{R_\nu}\rangle}{\langle {R_\nu}\rangle} = \alpha_T \cdot \frac{T_{eff}(t) - \langle{T_{eff}}\rangle}{\langle{T_{eff}}\rangle}
 \end{equation}
where $\langle{\dots}\rangle$ denote averages over the observation time and  $\alpha_T $ is the
effective correlation coefficient.

The  expected neutrino rate is given by the integration of the above defined production yield (Eq.~\ref{eq:prodyield}) and the effective detection area, over energy and atmospheric depth:
 \begin{equation} \label{eq:nurate}
      R_\nu (\theta) = \int \int \phi_N(E_\nu) \cdot  P(E_\nu, \theta, X ) \cdot A_{eff} (E_\nu , \theta)  \,dX\,dE_\nu ~.
 \end{equation}
The effective temperature is given by the expectation value of the atmospheric temperature profile weighted with the energy dependent production yield and the effective detection area: 
\begin{equation}
     \label{eq:teff}
     T_{eff} (\theta) \equiv
     { \int \phi_N(E_\nu) \cdot T (X, \theta) 
     \cdot A_{eff} (E_\nu,\theta)
     \cdot P(E_\nu,\theta,X) 
     \, dE_\nu \, dX
          \over
       \int  \phi_N(E_\nu) \cdot 
      A_{eff} (E_\nu,\theta)
     \cdot P(E_\nu,\theta,X) 
     \, dE_\nu \, dX   
     } ~.
\end{equation}
With these definitions, the linear correlation (Eq.~\ref{eq:corr}) between the relative neutrino rate and the temperature change is given by:
 \begin{equation} \label{eq:alpha}
     \alpha_T = \frac{T}{R_\nu} \frac{\partial R_\nu}{\partial T}~.
 \end{equation}
While the correlation of atmospheric muons with the atmospheric temperature has been studied in detail in~\cite{serap:2019icrc}, the statistics of atmospheric neutrinos have been insufficient to measure this effect precisely \cite{2013icrc}. Nevertheless, the atmospheric neutrino flux provides complementary information. Firstly, it probes the global atmospheric temperature. Secondly, relative to atmospheric muons, the relative contribution of kaons is larger \cite{Desiati:2010wt}.
Due to the choice of effective parameters, the integration of the steeply falling production yield with the steeply increasing effective area in Eq.~\ref{eq:teff} is dominated by the transition region between pion and kaon production. Hence, the effective temperature becomes sensitive to the relative production yield.
A further advantage of this type of correlation analysis is the use of the total observed rate. In contrast to energy-dependent spectra that are degraded by the poor experimental energy resolution and  affected by systematic uncertainties of the effective area, the total rate is statistically more significant. Note that in this approach the systematic uncertainties of the flux normalization and effective detector area largely cancel.

In this paper, we present for the first time a statistically significant measurement of this correlation using data from several years of operation of IceCube.

\section{Used Data sets}

\subsection{Atmospheric Temperatures}

%period 98.88 minutes
%orbits per day 14.5
%overpasses 2 per day: daytime ascending pass (south pole to north pole) at 1:30 pm local time; nighttime descending pass (north pole to south pole) occurs at 1:30 am local time
%daily coverage more than 95% of Earth’s surface
%swath width 1650 kilometers (1025 mi)
%resolution at nadir 13.5 kilometers (8.4 miles)
%resolution at scan extremes 41 x 21.4 kilometers (25.4 x 13.2 miles)
%samples per scanline across swath 90 footprints, each footprint is 1.1 degree
The atmospheric temperature profile is continuously and globally monitored by the  Atmospheric Infrared Sounder (AIRS) aboard NASA's Aqua satellite  \cite{aumann2003airs}, covering about \SI{95}{\percent} of the Earths surface every day.
\begin{figure}[htp]
	%\centering
	\floatbox[{\capbeside\thisfloatsetup{capbesideposition={right,bottom}, capbesidewidth=4.5cm}}]{figure}[\FBwidth]
	{\hspace{0.1cm}\caption{Level-3 temperature data in \si{K} from AIRS for the descending node at a pressure level of \SI{850}{hPa}}\label{fig:tempsky}}
	{\includegraphics[width=0.6\textwidth]{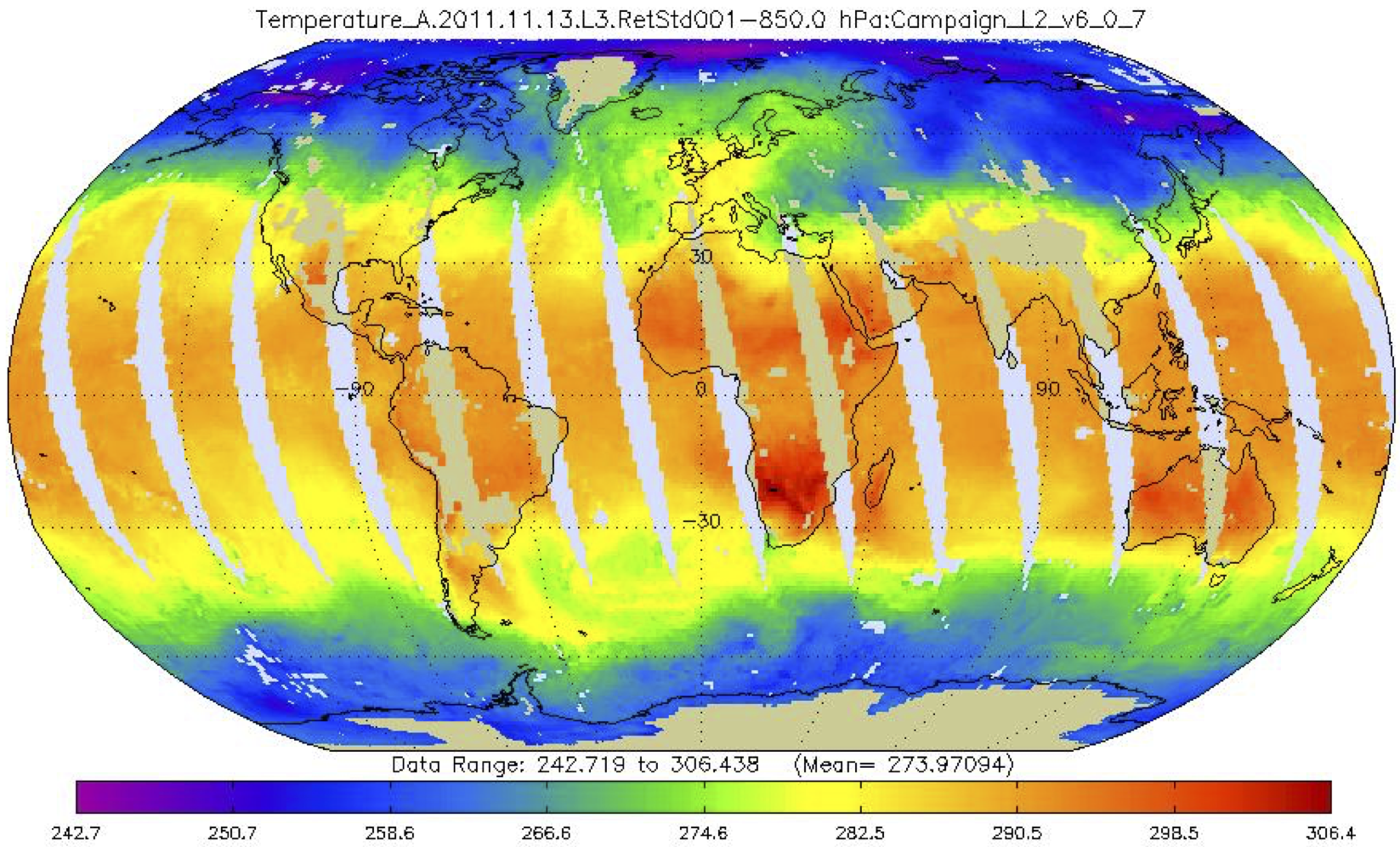}}
\end{figure}
The satellite orbits Earth \num{14.5} times per day on a highly inclined, nearly sun-synchronous polar orbit, on a northbound ascending orbit just after noon (1:30 p.m.) and a southbound descending orbit just after midnight (1:30 a.m.) local time.
The atmospheric temperature is measured by a pivotable infrared camera for \num{24} fixed pressure levels ranging from \SIrange{1.0}{1000}{hPa} and a ground swath of \SI{1650}{km} diameter.
Here, we use the level-3 data \cite{tian2013airs} that is binned in  $ {1}^{\circ} \times {1}^{\circ} $ cells in latitude and longitude.
An example data set is shown in  figure \ref{fig:tempsky}.
Data gaps that arise from e.g.\ non-perfect ground coverage at the equator are interpolated between neighbouring longitudes.
This fraction of  uncovered area becomes almost negligible for the polar zones. Additionally, we perform daily averages of the ascending and descending orbits.
Furthermore, there can be data loss at low altitudes due to rapid topography changes within one bin or local weather phenomena. Note that the neutrino production yield is small at low altitudes.

\begin{figure}[htp]
\centering
    \begin{minipage}{.75\linewidth}
    \centering
        \includegraphics[width=\linewidth]{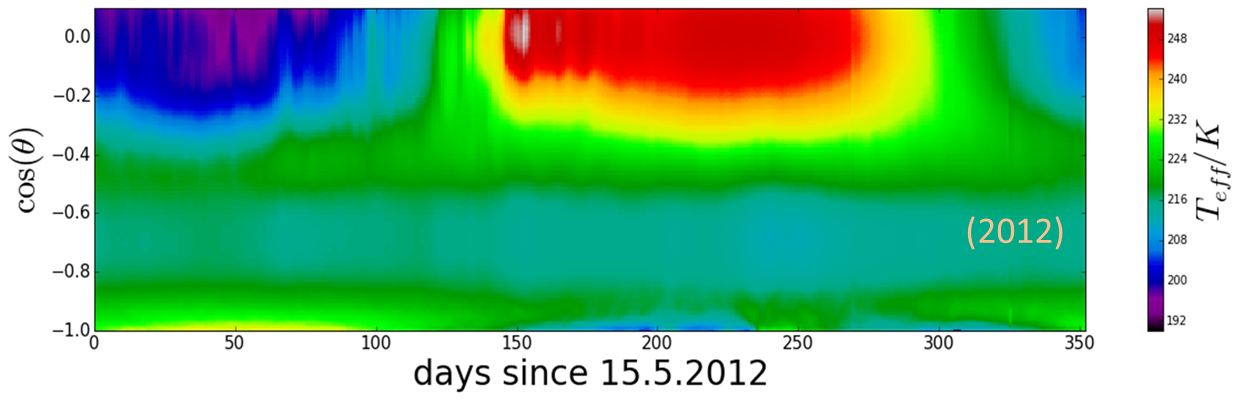}
    \end{minipage}
    \hfill
    \begin{minipage}{.75\linewidth}
    \centering
        \includegraphics[width=\linewidth]{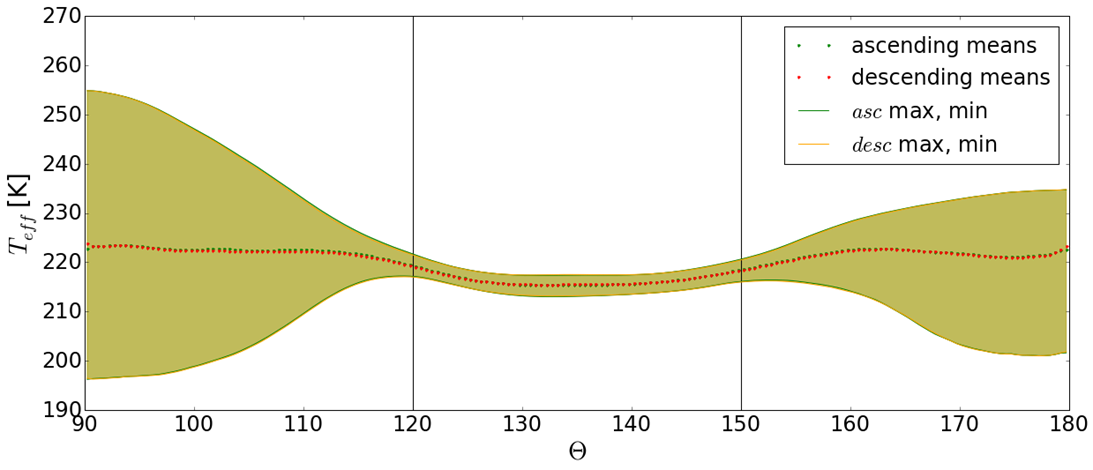}
    \end{minipage}
    \caption{Calculated values of $T_{eff}$ for about one year of AIRS level 3 data. Top panel (from \cite{jakobboettcher:bachelor}) shows $T_{eff} $ as function of time and observation zenith angle (in detector coordinates). Bottom panel (from \cite{fuerst:bachelor}) shows the median temperature (red line) versus observation zenith. Also shown as shaded regions are the \SI{5}{\percent} and \SI{95}{\percent} percentiles. Different line colors (hardly visible) represent data from the ascending and descending orbits and illustrate that no systematic difference is seen between the two. The vertical lines separate South, Equator and North region.}
	\label{fig:efftempsky}
\end{figure}

% Philipp Fürst page 23 
Figure \ref{fig:efftempsky} shows the calculated effective temperature for the period May 2012 to April 2013 as a function of the zenith angle of observed up-going neutrinos. The relation between latitude of the corresponding air shower and neutrino direction can be approximated as
$b = 2 \cdot(\theta  - \SI{135}{\degree}) $.
It can be seen that in the region $ \SI{120}{\degree} < \theta < \SI{150}{\degree} $, corresponding to equatorial latitudes $b$ between \SIrange{-30}{30}{\degree}, almost no variation ($<\SI{5}{K}$) is observed.
The Northern hemisphere region $\theta>\SI{150}{\degree}$ and the Southern hemisphere region $\theta< \SI{120}{\degree}$
are anti-correlated in their seasonal dependency.
The largest variation is observed for the Southern region. %The effective temperature of 
This region that also provides the majority of observed atmospheric neutrinos in IceCube
%(which also subtends the largest solid angle) 
is  considered in the following.

\subsection{Neutrino data}
For the neutrino data the same data set as used
for the measurement of astrophysical neutrinos \cite{Haack:2017dxi,Aartsen:2016xlq,d01:2019} is chosen. The data has been collected between May 2012 and May 2017, which represents a period of largely unchanged detector configuration. 
The data is based on a high  quality selection of well reconstructed up-going muons with a directional resolution $<\SI{1}{\degree}$ 
%resolution from rene paper
and $<\SI{0.3}{\percent}$ background contamination from atmospheric muons.
The data selection is unchanged for the full observation of this analysis. A high statistics of neutrino counts of typically \SI{140}{\per\day}
is observed in the South region.

\begin{figure}[htp]
\centering
	\floatbox[{\capbeside\thisfloatsetup{floatwidth=sidefil, capbesideposition={right, bottom}, capbesidewidth=4cm}}]{figure}%[\FBwidth]
	{\caption{Relative change of the daily measured neutrino rates and effective temperatures for the South region for the observation time of the analysis. The top panel shows the default \SI{1}{\day} binning, while the bottom panel shows the same data binned and averaged over \SI{30}{\day}. The colors in the top panel indicate different seasons.}\label{fig:nudata}}
	{\begin{minipage}[t]{.9\linewidth}
		\centering
	    \includegraphics[width=\linewidth]{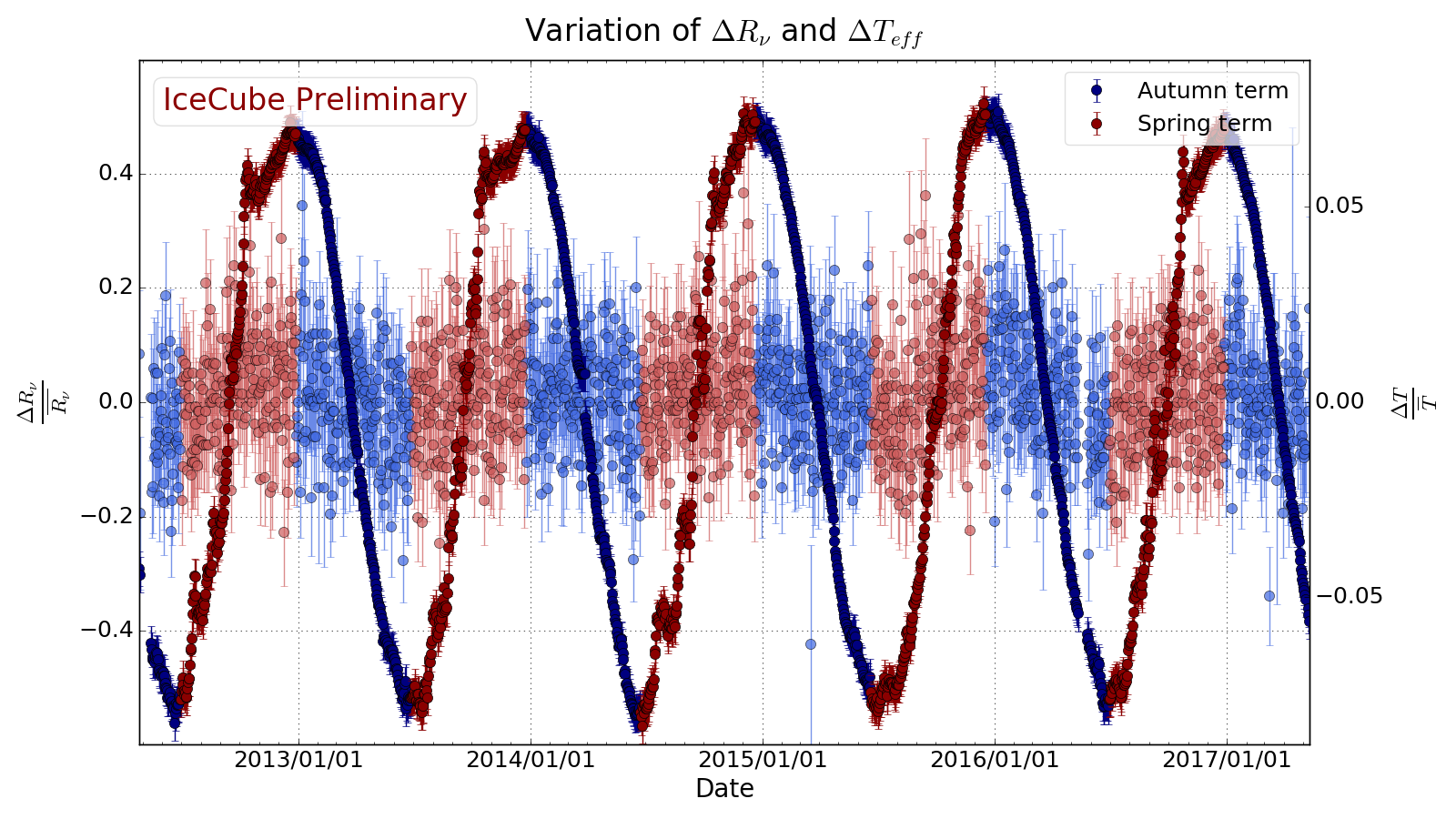}
	    \end{minipage}
	    %  <-- essential for both plots in the same line
	\hfill
	
	    \begin{minipage}[t]{.9\linewidth}
	    \centering
	    \includegraphics[width=\linewidth]{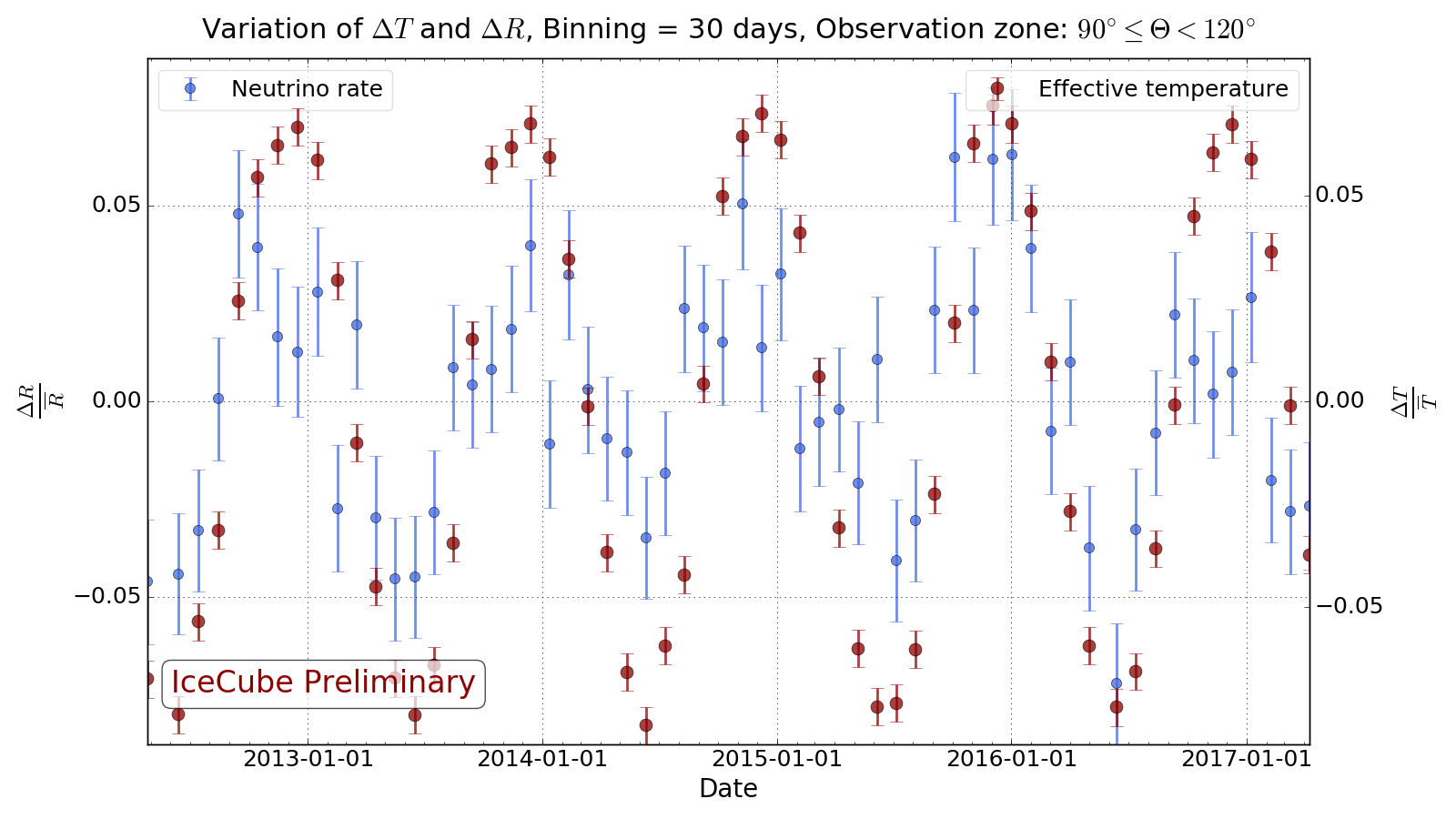}
	    \end{minipage}}
\end{figure}

The relative change of the daily measured neutrino rates, corrected for detector live time is shown in figure  \ref{fig:nudata} together with the calculated values of $T_{eff} $ in the South region.
Despite the large statistical fluctuation of the experimental rates, a correlated variation is indicated by the data. The correlation becomes obvious when the data is binned in intervals of \SI{30}{\day}.

\section{Results \label{sec:result}}

\begin{figure}[htp]
	\centering
		\begin{minipage}[b]{.49\linewidth}
		\centering
	    \includegraphics[width=\linewidth]{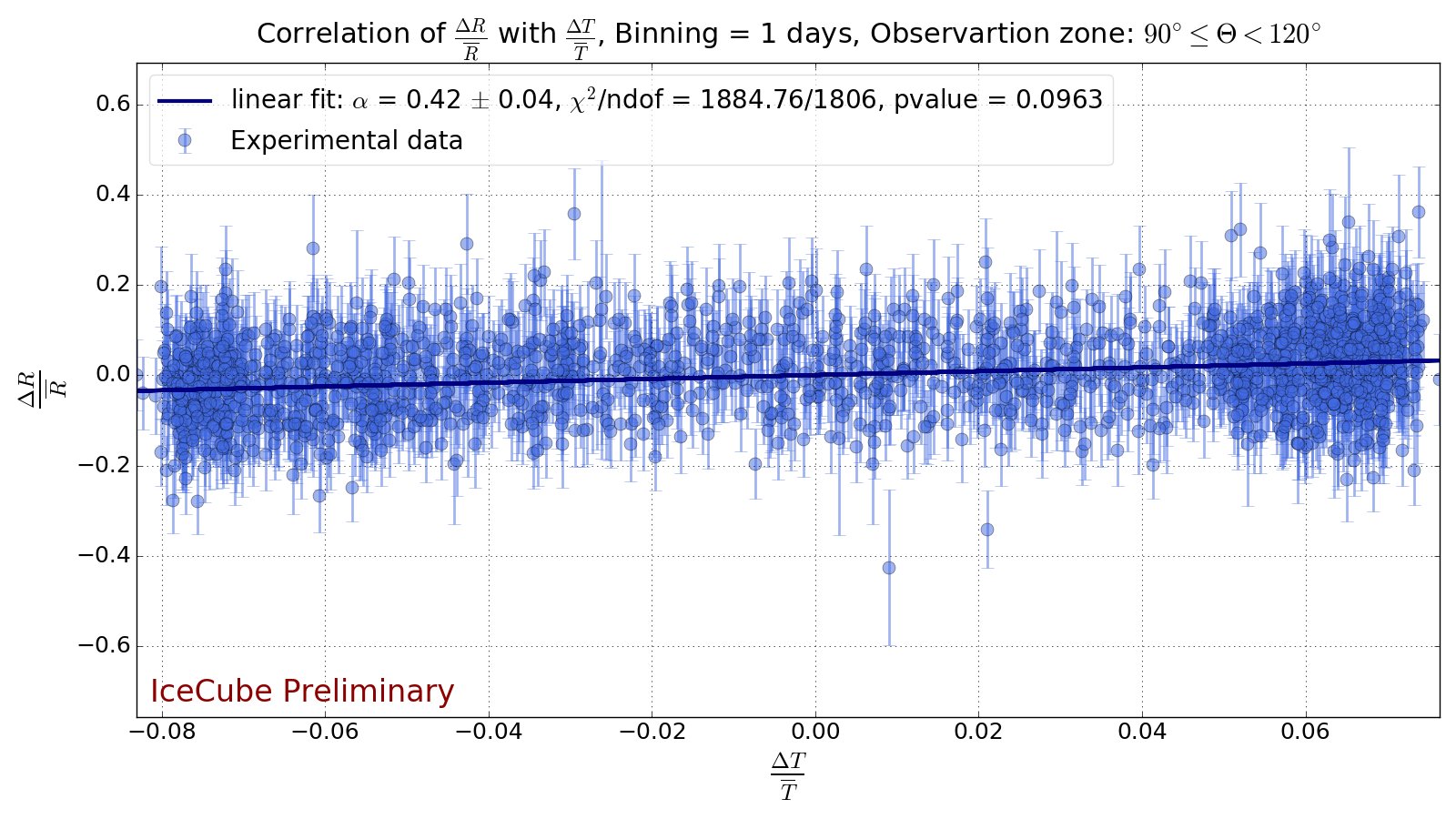}
	\end{minipage}%  <-- essential for both plots in the same line
	\hfill
	    \begin{minipage}[b]{.49\linewidth}
	    \centering
	    \includegraphics[width=1.\linewidth]{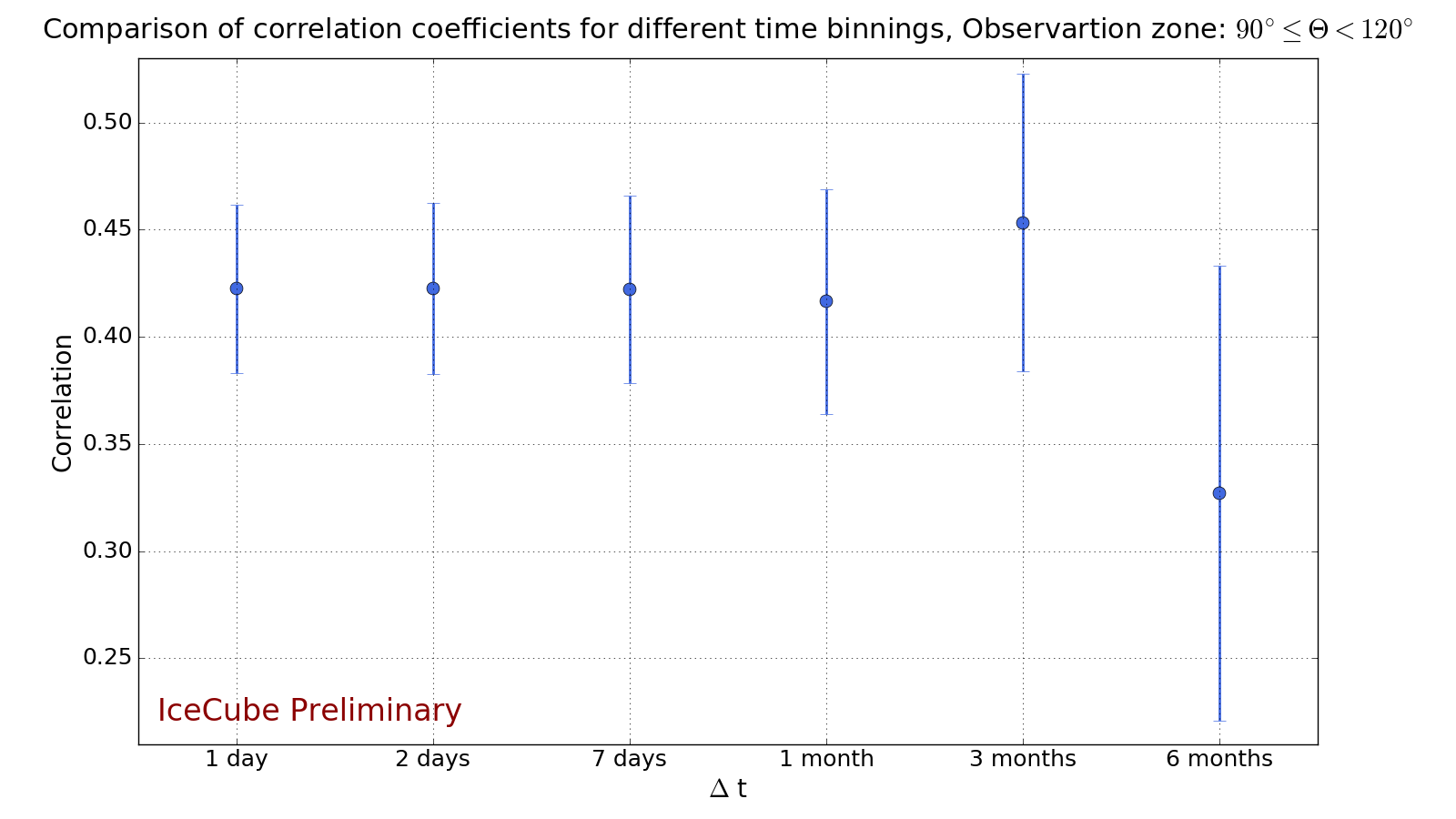}
	\end{minipage}%  <-- essential for both plots in the same line

	\caption{Correlation of atmospheric neutrino rate changes with relative changes of $T_{eff}$ (left panel) and comparison of the correlation coefficients for several time binnings (right panel).}
	\label{fig:fullresult}
\end{figure}

%\subsection{Yearly averaged result}
The correlation of relative temperatures and relative neutrino rates in the South zone for all observed days is shown in figure
\ref{fig:fullresult} (left panel). A linear fit results in a correlation coefficient of
\begin{equation}
    \alpha = 0.42 \pm 0.04
\end{equation}
with ${\chi}^{2} / ndof = 1884.76 / 1806$. The statistical significance of the correlation with respect to the non-correlation hypothesis $\alpha=0$ based on a $\chi^2$-test is \num{11} standard deviations.
This represents the most significant measurement to date of a seasonal modulation of atmospheric neutrinos.

In order to test the dependency of the result on the choice of binning, the analysis is repeated for larger time bins. Figure \ref{fig:fullresult}  shows that up to bin sizes of a month the result remains robust and is not affected by the choice of time bins. 

In order to test for systematic atmospheric effects in different seasons, the data is split into two subsets as illustrated by color in the top panel of figure \ref{fig:nudata}. The first set comprises the period January to June when the temperatures decrease after austral summer.
%The second set comprises dates from July to December when the atmospheric conditions are inverted and temperatures increase following austral winter. It can be seen in figure \ref{fig:nudata} that the temperature fluctuations are rather different in these two phases. In the period of rising temperatures, daily fluctuations are substantially stronger than for the phase of cooling atmosphere.
The second set comprises dates from July to December when the atmospheric conditions are inverted, and daily temperature fluctuations are also substantially stronger.
%\begin{figure}[h]
% 	\begin{minipage}[b]{.49\textwidth} \centering
% 	\includegraphics[width=\textwidth]{Plots/Teff_seasonsmarked.png}\\ 
% 	\end{minipage}%  <-- essential for both plots in the same line
% 	\hfill
% 	\begin{minipage}[b]{.49\textwidth} \centering
% 	\includegraphics[width=\textwidth]{Plots/DelRdRdelTdT_seasonsmarked.png}\\
% 	\end{minipage}
	
%  	\begin{minipage}[b]{.49\textwidth}
%  	\centering
%	\includegraphics[width=\textwidth]{Plots/Alphas_falling_withoutpulls.png}
%	\end{minipage}%  <-- essential for both plots in the same line
%	\hfill
%	\begin{minipage}[b]{.49\textwidth}
%	\centering
%	\includegraphics[width=\textwidth]{%Plots/Alphas_rising_withoutpulls.png}
%	\end{minipage}
  
%    \caption{Fits of different seasonal temperature gradients
%    Left: Selected data Right: Fits of the two seasons
%    Fitted values are $\alpha^- =0.45 \pm 0.06$	and $ \alpha^+ =0.40 \pm 0.06$
%     Left: (A) Season of falling temperatures Right: (B) Season of rising temperatures}

%     \label{fig:season1}
    
%\end{figure}

%In figure \ref{fig:season1} 
The two data sets are fitted separately for the correlation coefficient resulting in {${\alpha}^{-} = 0.45 \pm 0.06$} and {${\alpha}^{+} = 0.40 \pm 0.06$}
%\begin{equation}
%\alpha^- =0.45 \pm 0.06 \qquad \mbox{and} \qquad  \alpha^+ =0.40 \pm 0.06
% \end{equation}
where $+$ denotes increasing temperatures and $-$ decreasing temperatures respectively.
The two results are statistically consistent at the \SI{1}{\sigma} level.
Yet,  the data set of stronger temperature fluctuations shows a smaller correlation, indicating  that averaging the satellite data during these more fluctuating periods may require finer binning in time and direction.
However, given the current uncertainties, such a systematic effect cannot be tested for.

\section{Systematic uncertainties and comparison with  expectation}

For the discussion of systematics, uncertainties in relative temperatures and relative neutrino rates and the averaging of data over the South zone have to be taken into account.
%The uncertainties of the temperature data  are discussed above and amount to about \SI{1}{\percent} in relative temperature.
The accuracy of the calculation of $T_{eff}$ is limited by the accuracy of the measured temperatures which are estimated \cite{olsen2017airs} as \SI{1}{K/km}.
Assuming additionally a height uncertainty of \SI{1}{\percent} in the integration of the atmospheric depth. This implies an uncertainty of $T_{eff} $ of \SI{0.25}{\percent}.
The standard deviation of the difference of the ascending and descending temperatures corresponds to \SI{0.45}{K}. This translates into an uncertainty of the relative temperature of \SI{0.2}{\percent}.
As discussed above, fractional ground coverage and topographic  changes ($<\SI{0.1}{\percent}$)
can be neglected for the South region \cite{backes:bachelor}.
Most relevant is the uncertainty of the numerical  numerical integration of slant depth. Depending on the choice of bin-boundaries the relative temperatures can change up to $\SI{1}{\percent}$ \cite{fuerst:bachelor}.
Here we centre the bins around the given altitude levels, i.e. letting the up-most bin extend to $X=0$.

For the measurement of neutrino rates, uncertainties in the relative live time for each day of operation and background from wrongly reconstructed atmospheric muons have to be considered. 
The background of atmospheric muons amounts to about \SI{0.4}{\per \day}. It is  expected to be correlated with the same seasonal phase as the neutrinos in the South zone but with a correlation factor twice as large. Assuming relative temperature changes up to \SI{8}{\percent} this results in a negligible variation of the rate at maximum of $ \Delta R \approx \SI{0.06}{\per \day}$.
The uncertainty in the calculation of live time can be neglected as IceCube operates close to \SI{99}{\percent} uptime and the downtime can be calculated to an accuracy of a few \si{ms}.

\begin{figure}[htp]
\centering
	\floatbox[{\capbeside\thisfloatsetup{capbesideposition={right, bottom}, capbesidewidth=5cm}}]{figure}[\FBwidth]
	{\caption{Calculated $\alpha$ for the standard US atmosphere and air shower model (I)}\label{fig:alphacalc}}
	{\includegraphics[width=0.45\textwidth]{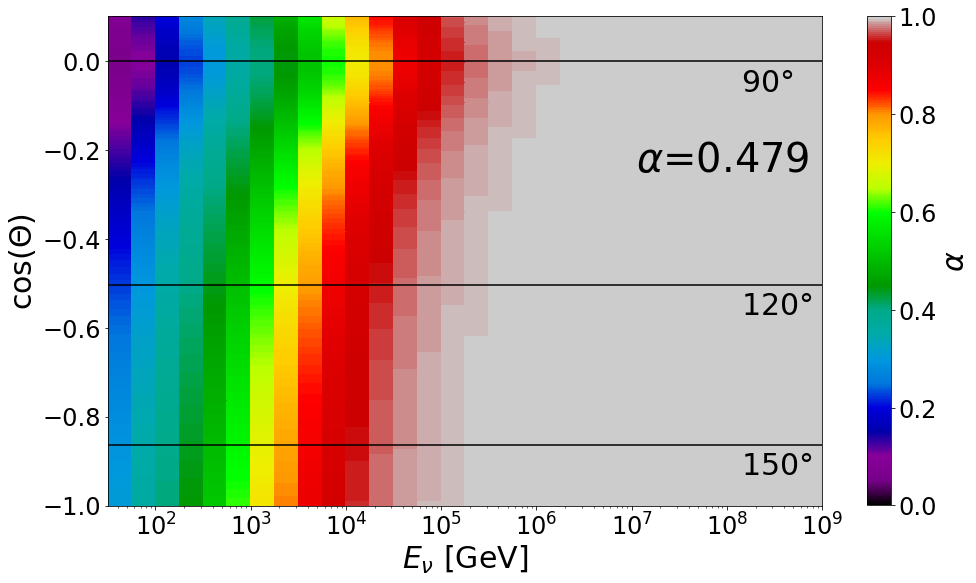}}
\end{figure}

The value of $\alpha $ depends on the zenith angle. Calculations based on the simple approximation Eq.~\ref{eq:prodyield} indicate a maximum difference of  $\alpha (\ang{120}) - \alpha(\ang{90}) \approx 0.15 $ corresponding to one standard deviation of $\sigma(\alpha) = \pm 0.04 $ within the observation region (see also figure \ref{fig:alphacalc}). This is of similar order as the statistical uncertainty but does bias of the average result.
In summary, systematic uncertainties are on the level of \SI{1}{\percent} and thus about a factor 10 smaller than the statistical uncertainty.

%\href{https://wiki.icecube.wisc.edu/index.php/Neutrino_Meteorology}{Marit}
%\href{https://wiki.icecube.wisc.edu/index.php/Prediction_of_seasonal_variations_of_atmospheric_neutrinos}{Patrik}
For an estimation of the theoretical expectation the above approximation Eq.~\ref{eq:prodyield}
is inserted into the definition of $\alpha$ (Eq.~\ref{eq:alpha}),
and the resulting $\alpha$ is calculated using the standard US atmosphere \cite{united1976us}.
For the calculation, different values for the spectrum weighted moments (Z-factors) and atmospheric interaction lengths ($\Lambda$-values) are used as given in 
\cite{Gaisser:2016uoy}.
The following models are
tested: (I) Sibyll 2.3 \cite{Riehn:2015oba} using constant Z-factors and energy dependent $\Lambda$-values, (II) using constant values of Z  as provided by \cite{Sanuki:2006yd} for \cite{Gaisser:2016uoy} and constant $\Lambda$-values based on Sibyll 2.1 \cite{Ahn:2009wx} calculations, and (III) constant values of Z-factors from \cite{Gaisser:1990vg} and the same $\Lambda$-values as (II).

An example of the resulting $\alpha $ as a function of energy and zenith angle is shown in figure \ref{fig:alphacalc}.
All the values for $\alpha$ averaged over the South observation zone and energy are given in table \ref{tab:theoryresult}.

\begin{figure}[htp]
\centering
	\floatbox[{\capbeside\thisfloatsetup{capbesideposition={right, bottom}, capbesidewidth=5cm}}]{table}[\FBwidth]
	{\hspace{2cm}\caption{Theoretical estimates of the expected correlation}\label{tab:theoryresult}}
	{\begin{tabular}{c|c|c|c}
       Model  &  (I) & (II) & (III)\\
       \hline
        $\alpha$  & \num{0.479}& \num{0.496} & \num{0.484}\\
    \end{tabular}}
\end{figure}

These expectations differ from the experimental measurement by \numrange{1.5}{2} standard deviations. 
This tension may hint at a higher than expected contribution of kaons to the production of atmospheric neutrinos.
However, the large uncertainty of the simplified approximation has also to be taken into account.
The variation in $\alpha$ with changes of different parameters is up to $0.02$. 
The effect of using the US atmospheric model for the South Pole atmosphere is of similar size.

\section{Conclusion and Outlook\label{sec:conclusion}}

The measurement of the temperature dependent variation of the flux of atmospheric neutrinos is highly interesting, because 
such a measurement, being independent of the total flux normalization, can constrain hadronic interaction models of atmospheric air showers, in particular the relative fraction of kaons and pions.

The analysis presented here has established for the first time a highly significant correlation of the rate of atmospheric neutrinos with the atmospheric temperature. 
The measured correlation of $\alpha = 0.42 \pm 0.04$ is in reasonable agreement with the expectation of a simplified  approximation \cite{Gaisser:2016uoy} of $\alpha = 0.49 \pm 0.03$. The somewhat lower measured value may hint at an increased kaon fraction in atmospheric air showers.
More precise calculations using the full cascade equations governing atmospheric air showers are currently being performed with MCEq \cite{Fedynitch:2015zma} for a more precise prediction of the expected correlation and the future test of hadronic interaction models.

The experimental analysis can also be substantially improved, because it neglects variations of the atmosphere within the considered zenith range and variations in longitude.
Examples of such longitudinal variations are Rossby waves \cite{rhines2003rossby}. Rossby waves are regions of different temperatures rotating around the South Pole region, and are seen in the time-dependent AIRS temperature data.
Furthermore, the value of $\alpha$  is expected to considerably vary within  the considered zenith range.
This can be addressed well by an unbinned likelihood method, that naturally accounts for the full directional temperature information of each observed neutrino direction and is thus expected to further improve the accuracy and significance of the measurement in the future.

%\section{Listing some References}\label{sec:refs}

%Bla

%This is a paper from a previous ICRC %\cite{Zoll:2015wcu}. 
%This is a second paper from a previous ICRC \cite{Peiffer:2017vsm}. 
%This is a paper from the current ICRC \cite{Hussain:2019icrc-gw}.
%Here is an IceCube journal paper \cite{Aartsen:2016nxy} and an external journal paper \cite{Waxman:1998yy}.

% Set up the bibliography using BibTeX.
% Get references from inspirehep.net or NASA/ADS and put them in references.bib.
\bibliographystyle{ICRC}
\bibliography{references}

% Or, set up the bibliography manually, if you prefer to do things this way.
%
% \begin{thebibliography}{99}
%   \bibitem{Zoll:2015wcu}{{\bf IceCube} Collaboration, \pos{PoS(ICRC2015)1099} (2016).}
%   \bibitem{Peiffer:2017vsm}{{\bf IceCube-Gen2} Collaboration, \pos{PoS(ICRC2017)1052} (2018).}
%   \bibitem{Hussain:2019icrc_gw}{{\bf IceCube} Collaboration, \pos{PoS(ICRC2019)xyz} (these proceedings).}
%   \bibitem{Aartsen:2016nxy}{{\bf IceCube} Collaboration, M.~G.~Aartsen {et al.}, \emph{JINST} {\bf 12} (2017) P03012%
%   % optionally add arXiv ID here [{\tt astro-ph/1612.05093}]
%   .}
%   \bibitem{Waxman:1998yy}{E. Waxman and J. N. Bahcall, \emph{Phys. Rev.} {\bf D59} (1999) 023002.}
% \end{thebibliography}

\end{document}